\newcommand{\beq}{\begin{equation}}
\newcommand{\eeq}{\end{equation}}
\newcommand{\ber}{\begin{eqnarray}}
\newcommand{\eer}{\end{eqnarray}}
\documentclass{ws-ijmpd}

\begin{document}

\markboth{S Arbabi Bidgoli, M S Movahed and S Rahvar} {A
Parameterized Variable Dark Energy Model }

%
\catchline{}{}{}{}{}
%

\title{A Parameterized Variable Dark Energy Model: Structure Formation
and Observational Constraints  }

\author{S Arbabi Bidgoli}

\address{Institute for Studies in Theoretical Physics and Mathematics,
P.O.Box 19395--5531, Tehran, Iran \\
arbabi@ipm.ir}

\author{M S Movahed}

\address{Institute for Studies in Theoretical Physics and Mathematics,
P.O.Box 19395--5531, Tehran, Iran \\
Department of Physics, Sharif University of Technology,
P.O.Box 11365--9161, Tehran, Iran\\
m.s.movahed@mehr.sharif.edu}

\author{ S Rahvar}

\address{Institute for Studies in Theoretical Physics and Mathematics,
P.O.Box 19395--5531, Tehran, Iran \\
Department of Physics, Sharif University of Technology,
P.O.Box 11365--9161, Tehran, Iran\\
rahvar@sharif.edu}

\maketitle

\begin{history}
\received{Day Month Year}
\revised{Day Month Year}
\comby{Managing Editor}
\end{history}

\begin{abstract}
In this paper we investigate a simple parameterization scheme of the
quintessence model given by Wetterich (2004). The crucial parameter
of this model is the bending parameter $b$, which is related to the
amount of dark energy in the early universe. Using the linear
perturbation and the spherical infall approximations, we investigate
the evolution of matter density perturbations in the variable dark
energy model, and obtain an analytical expression for the growth
index $f$. We show that increasing $b$ leads to less growth of the
density contrast $\delta$, and also decreases the growth index.
Giving a fitting formula for the growth index at the present time we
verify that the approximation relation $f\simeq\Omega_m^{\alpha}$
also holds in this model. To compare predictions of the model with
observations, we use the Supernovae type Ia (SNIa) Gold Sample and
the parameters of the large scale structure determined by the
$2$-degree Field Galaxy Redshift Survey ($2$dFGRS). The best fit
values for the model parameters by marginalizing on the remained
ones, are $\Omega_m=0.21_{-0.06}^{+0.07}$,
$w_0=-2.05_{-2.05}^{+0.65}$ and $b=4.05^{+7.05}_{-2.25}$ at
$1\sigma$ confidence level. As a final test we calculate the age of
universe for different choices of the free parameters in this model
and compare it with the age of old stars and some high redshift
objects. Then we show that the predictions of this variable dark
energy model are consistent with the age observation of old star and
can solve the "age crisis" problem.
\end{abstract}


\section{Introduction}

Recent observations of type Ia Supernovae (SNIa) at low and medium
redshifts and the cosmic microwave background (CMB) anisotropies
strongly indicate that the total matter-energy density of the
universe is now dominated by dark energy
\cite{rie+al98,Riess00,per+al99,R04}. The origin and nature of this
dark energy term remains unknown. The most obvious theoretical
candidate of dark energy is the cosmological constant $\Lambda$
which has the equation of state $w=-1$ \cite{zel67,weinberg2,car}.
Since the cosmological constant is a physical link between micro and
macro scales, two main questions arise. First, the fine-tuning
problem asks why the dark energy density today is so small compared
to typical particle scales. The dark energy density is of order
$10^{-47}$ GeV$^{4}$, which appears to require the introduction of
new mass scale $14$ or so orders of magnitude smaller than the
electro-weak scale. The second difficulty, the cosmic coincidence
problem, states, Since the energy densities of dark energy and dark
matter scale so differently during the expansion of universe why are
they nearly equal today? To get this coincidence, it appears that
their ratio must be set to a specific, infinitesimal value in the
very early universe. As an attempt to solve these problems, one can
consider a scalar field with a potential and kinetic term which has
an equation of state to behave as vacuum energy. The energy density
of this field, called quintessence in the cosmological attractor
solution follows the energy density of matter and radiation, but
remains negligible until recent epoch \cite{CDS99}. In these models
the standard cosmological constant $\Lambda$-term is replaced by a
dynamical, time-dependent component. Many different variable dark
energy models have been proposed in the literature, e.g., Wetterich
(1988), Ratra $\&$ Peebles (1988) \cite{W88,RP88}. For a more
complete list of references and a review of this topic see
\cite{S05}. This is still not a satisfactory physical explanation
for the observed values of the cosmological constant term. The
quintessence models do not offer a fundamental explanation, instead
they are a phenomenological approach to express our inability of
understanding the nature of the cosmological constant in terms of a
variable scalar field.

Here we examine a generic parameterization of quintessence models
given by Wetterich (2004) \cite{W04}. In the recent paper
\cite{Doran05} some observational constraints have been investigated
for this model and the value of present equation of state was
assumed $w_0\geq-1$ but in this paper we let this parameter to be
$w_0\leq 0$ and give the best fitting values for $w_0$, present
matter density and bending parameter. Also in \cite{sadegh06} the
constraints related to the background evolution have been used. Our
motivation for the present work is to see the possible observational
effects \cite{all04,DD03,DD04} of a variable dark energy on the
growth of the large scale structure of the universe. To put the
rigorous constraints on the parameters of variable dark energy model
we use the luminosity distance of Supernovae type Ia (SNIa) of the
Gold Sample \cite{R04} and large scale structure parameters
determined by the $2$dFGRS team. The equation of state used for this
quintessence model is:
 \beq w(z;b,w_0)=\frac{w_0}{[1+b\ln(1+z)]^2},
\label{eq1} \eeq
where $w_0$ is the state parameter at the present
time and $b$ is the bending parameter, which expresses the change in
the equation of state of dark energy with redshift and is related to
the amount of dark energy in the early universe. According to the
theory of structure formation, the evolution of structure in the
linear and non-linear regimes depends strongly on the background
dynamics of universe. Dark energy as a crucial element of cosmic
fluid affects the dynamics of the universe and consequently changes
the growing rate of structure. By increasing the parameter $b$ in
the model, the universe enters earlier in the phase of dark energy
domination and the faster dilution of matter suppresses the
formation of further structure. We assume that variable dark energy
behaves as a smooth component, so that in our analysis the structure
formation is only due to matter condensation, while the variable
dark energy alters only the background cosmic dynamical evolution.
We mostly work with $\Omega_{tot}=1$ which is based on the results
of CMB experiments \cite{sp103}.

This paper is organized as follows. In section 2 we discuss the
linear perturbation theory of cosmological structure formation
applied to the case of a variable dark energy. Considering the
parameters of the model, we compute the evolution of the matter
density contrast and the evolution of the growth index. We apply the
spherical approximation with different initial conditions for
overdense and underdense regions and compare these results with the
linear approximation in Section 3. In section 4, we constrain the
parameters of model by using high redshift Supernovae type Ia (SNIa)
of the Gold sample and the parameters of Large Scale Structure (LSS)
determined by $2$-degree Field Galaxy Redshift Survey ($2$dFGRS). As
a consistency test we look at the predicted age of universe in the
model by comparing with the age of high redshift objects. The
conclusions are given in Section 5.

\section{Linear Newtonian structure formation with variable dark energy}

\begin{figure}[pb]
\centerline{\psfig{file=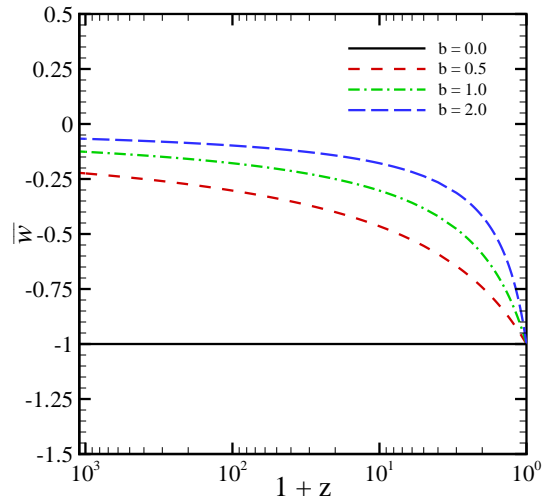,width=8cm}} \vspace*{8pt}
\caption{$\bar{w}(z;b,w_0)$ in terms of redshift for various bending
parameters. Here we choose $w_0 =-1$. The slope of the graphs are
more sensitive to the bending parameter at low redshifts rather than
high redshifts. \label{1}}
\end{figure}

\begin{figure}[pb]
\centerline{\psfig{file=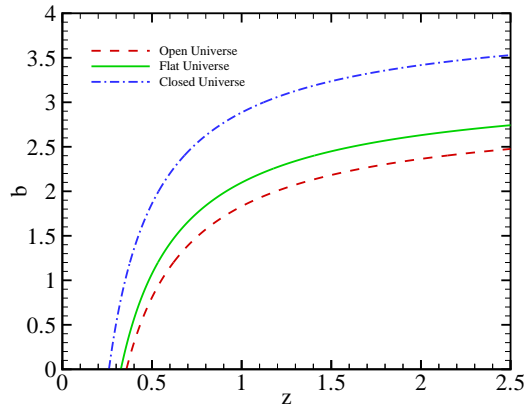,width=8cm}} \vspace*{8pt}
\caption{Redshift of equality of dark energy and dark matter density
as a function of the bending parameter $b$, for a flat universe
(Solid line) with the $\Omega_{m}=0.3$ and $\Omega_{\lambda}=0.7$,
an open universe (dashed line) with $\Omega_{m}=0.2$ and
$\Omega_{\lambda}=0.5$ and a closed universe (dot-dashed line) with
$\Omega_{m}=1.0$ and $\Omega_{\lambda}=2.0$. Increasing the bending
parameter results in a higher redshift of equality. The present
state parameter is set to $w_0=-1.0$. \label{2}}
\end{figure}

The dynamics of the universe is driven by the Friedmann's equations
as:
\begin{eqnarray} &&\left(\frac{\dot a}{a}\right)^2+\frac{k}{a^2}=\frac{8\pi
G}{3}\left(\rho_m+\rho_{\lambda}\right),\\
&&\dot{\rho}_m+3\frac{\dot{a}}{a}\rho_m=0,\label{c1}\\
&&\dot{\rho}_{\lambda}+3\frac{\dot{a}}{a}\rho_{\lambda}[1+w(a;b,w_0)]=0,\label{c2}
\end{eqnarray}
where the dot denotes the derivative with respect to time, $\rho_m$
and $\rho_{\lambda}$ stand for the pressureless matter and
quintessence component, respectively. Also $w(a;b,w_0)$ is given by
Eq. (\ref{eq1}). As usual, $k=0,1,-1$ indicates a flat, closed and
open spacial section. Dynamic equations for each fluid (equations
(\ref{c1}) and (\ref{c2})) are: \beq
\rho_m=\rho_m^0a^{-3}\qquad,\qquad
\rho_{\lambda}=\rho_{\lambda}^0a^{-3(1+\bar{w}(a;b,w_0))},\eeq where
$\bar{w}(z;b,w_0)=w_0/[1+b\ln(1+z)]$. Fig. \ref{1} shows the
behavior of $\bar {w}(z;b,w_0)$ as a function of redshift. We notice
that most variation in $\bar{w}$ takes place at low redshifts and
that larger values of $b$ correspond to a more dust-like behavior of
dark energy at high redshifts. The redshift of equality of matter
and dark energy strongly depends on the choice of parameter $b$, as
shown in Fig. \ref{2}, a universe with a large $b$ enters the dark
energy dominated regime earlier.

The Hubble parameter for a universe composed of dark matter and dark
energy is given by: \beq
H^2(z;b,w_0)=H_0^2[\Omega_m(1+z)^3+\Omega_\Lambda(z;b,w_0)-(\Omega_{tot}-1)(1+z)^2],
\label{eq4} \eeq with the definitions:
 \beq \Omega_m=\frac{8\pi
G}{3}\frac{\rho_m^0}{H^2_0},\eeq \beq \Omega_\Lambda(z;b,w_0)=
\frac{8\pi
G}{3}\frac{\rho_{\lambda}^0}{H^2_0}(1+z)^{3[1+\bar{w}(z;b,w_0)]},
\label{eq11} \eeq \beq
\Omega_{tot}=1-\frac{k}{H^2_0}=\Omega_m+\Omega_{\lambda},\eeq where
$\Omega_{\lambda}$ and $\Omega_{m}$ are the ratio of the dark energy
density and dark matter density to the critical energy density at
the present time, respectively.

\begin{figure}[pb]
\centerline{\psfig{file=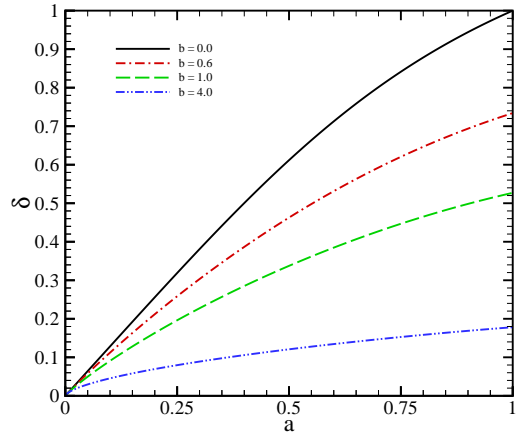,width=8cm}} \vspace*{8pt}
\caption{Evolution of density contrast as a function of scale factor
for different bending parameters in the flat universe with $\Omega_m
= 0.3$, $\Omega_{\lambda}=0.7$ and $w_0=-1.0$. \label{3}}
\end{figure}

Using the continuity and Poisson equations in the expanding FRW
universe, the evolution of density contrast, $\delta=\delta
\rho/\bar{\rho}$ in the linear approximation is given by
\cite{P93,B04}:

\beq \ddot{\delta}+2\frac{\dot{a}}{a}\dot{\delta} - \left( {v_s}^2
\nabla^2 +4\pi G \rho_m\right) \delta=0. \label{eq2} \eeq  In Eq.
(\ref{eq2}) the dark energy enters through its influence on the
expansion rate $H(a;b,w_0)$. The validity of this linear Newtonian
approach is restricted to perturbations on the subhorizon scales but
large enough where structure formation is still in its linear phase
\cite{P93,B04}. We also assume that the sound horizon of dark energy
is much larger than the wavelength of the perturbations, so we do
not need to consider the clustering of dark energy. If the
perturbation is larger than the Jeans length, $ \lambda_J= \pi^{1/2}
v_s /  \sqrt{G \rho_m} $, then Eq. (\ref{eq2}) for cold dark matter
(CDM) density contrast reduces to:
 \beq \ddot{\delta}+2
\frac{\dot{a}}{a} \dot{\delta}-4\pi G \rho_m\delta=0, \label{eq3}
\eeq The equation that describes the evolution of density contrast
with respect to the scale factor is:

\beq
\frac{d^2\delta}{da^2}+\frac{d\delta}{da}\left[\frac{\ddot{a}}{\dot{a}^2}+\frac{2H(a;b,w_0)}{\dot{a}}\right]-
\frac{3H_0^2}{2\dot{a}^2a^3}\Omega_{m}\delta=0. \label{eq31} \eeq
The numerical solution of Eq. (\ref{eq31}) in a Friedmann universe
with variable dark energy evolving as given in Eq. (\ref{eq4}), is
shown in Fig. \ref{3}. In the matter dominated era, the density
contrast $\delta$ grows linearly with the scale factor, while we
have a deviation from a linear behavior, when dark energy begins to
dominate. For larger $b$ (see Fig. \ref{2}) the domination of dark
energy occurs earlier and the friction term in Eq. (\ref{eq3})
increase so the growth of density contrast decreases.
The relative difference between the present value of the
$\Lambda$CDM and the dark energy models is about $2.5\%$ for $b=0.1$
and $100\%$ for $b=1$.

In the linear perturbation theory the peculiar velocity field
$\bf{v}$ is determined by the density contrast \cite{P93,P80} :

\beq {\bf v} ({\bf x})= H_0 \frac{f}{4\pi} \int \delta ({\bf y})
\frac{{\bf x}-{\bf y}}{\left| {\bf x}-{\bf y} \right|^3} d^3 {\bf
y}, \eeq where the growth index $f$ is defined as:

\beq f=\frac{d \ln \delta}{d \ln a}. \label{eq5} \eeq

We use the density contrast $\delta$ to compute the growth index of
structure $f$, which is an important quantity for the interpretation
of peculiar velocities of galaxies, as discussed in \cite{P80} for
the Newtonian and \cite{rah02} for the relativistic regime of
structure formation. For understanding the physical meaning of the
growth index it is helpful to divide the second term of Eq.
(\ref{eq3}) (friction) by the third (Poisson), which shows that: $f
\propto 2H\dot{\delta}/4\pi G \rho_m\delta$.

\begin{figure}[pb]
\centerline{\psfig{file=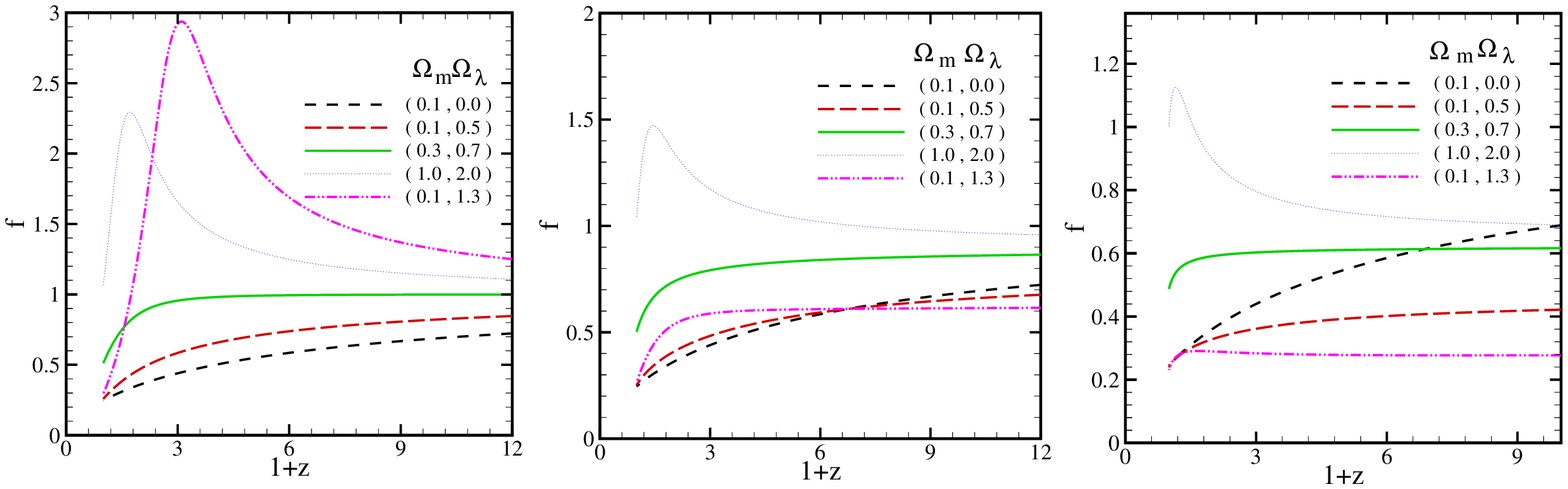,width=15cm}} \vspace*{8pt}
\caption{Growth index versus redshift for pairs of cosmological
density parameters $\Omega_m$ and $\Omega_{\lambda}$ The panels
refer to different values of the bending parameter $b$ and
$w_0=-1.0$. Left panel: $b=0.0$ corresponds to the case of
$\Lambda$CDM model. Middle: $b=1.0$. Right: $b=4.0$ as an extreme
case of the variable dark energy model. \label{4}}
\end{figure}

According to equations (\ref{eq31}) and (\ref{eq5}), the evolution
of the growth index is given by:

\begin{eqnarray}
\label{index} &&\frac{df}{d\ln
a}=-f\left[2-\frac{H_0^2}{2}[\frac{2}{H_0^2}+\frac{\Omega_m}{a^3}+\Omega_{\Lambda}(a;b,w_0)(1+3w(a;b,w_0))]
\right] \nonumber\\ && \qquad\qquad
-f^2+\frac{3H_0^2}{2a^3}\Omega_m.
\end{eqnarray} In Fig. \ref{4} we plot the numerical
solution of Eq. (\ref{index}) for the growth index as a function of
redshift for $b=0$, $b=1.0$ and $b=4.0$, for various amounts of dark
matter and dark energy at the present epoch. The growth index $f$
has been calculated for models with a cosmological constant earlier
and the case of $b=0$ corresponds to the $\Lambda$CDM model with
non-variable cosmological constant, where our results are identical
with the earlier results given by \cite{LLPR91}. There a fitting
formula was given for the dependence of $f$ on $\Omega_m$ for the
case of a flat universe. The dependence on the cosmological constant
was found to be very weak. Here we extend this analysis to models
with a variable dark energy, including the bending parameter $b$ and
$w_0$. In agreement with the tendency found earlier, we also find
that in our case the dependence on $b$ is not significant at the
present time. This fitting formula is as follows:
\begin{eqnarray}
f(\Omega_m,\Omega_{\lambda},b,w_0,z=0)& \approx &
\Omega_{m}^{0.57}+[-0.030+0.019\exp(-b)-0.038w_0] \nonumber\\ &
\times &
\Omega_{\lambda}^{[1.047-0.145b^{0.51}+(-0.104-0.30b^{1.2})w_0^3]}-\frac{b^{0.45}}{100}.
\label{fit}\end{eqnarray}

\begin{figure}[pb]
\centerline{\psfig{file=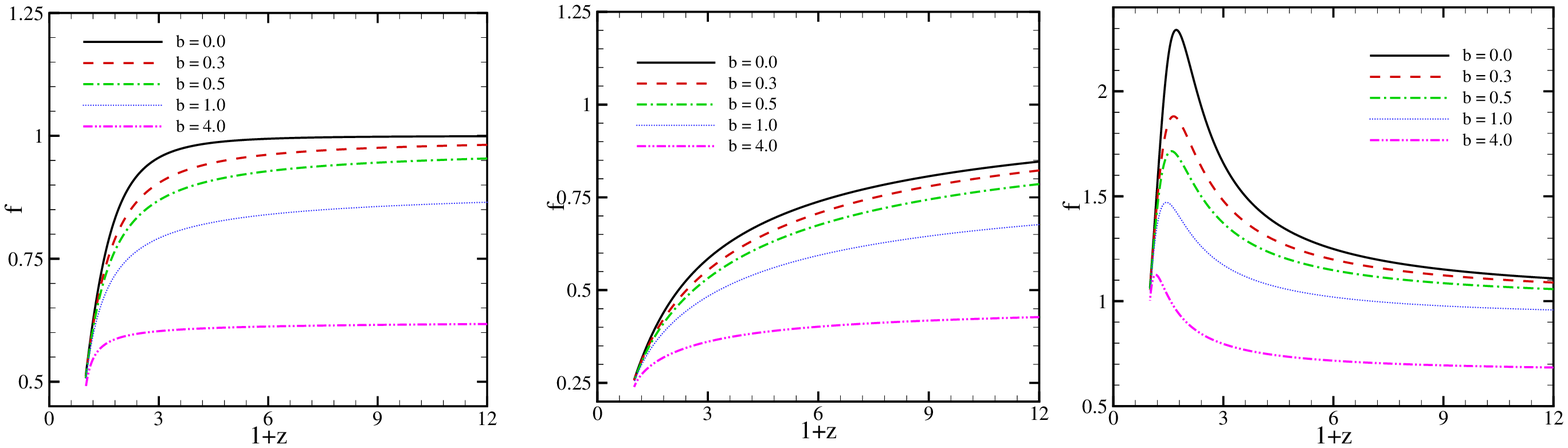,width=15cm}} \vspace*{8pt}
\caption{The effect of bending parameter $b$ on the evolution of the
growth index as a function of redshift in case of different pairs of
$\Omega_m$ and $\Omega_{\lambda}$. Left panel: A flat universe with
$\Omega_{m}=0.3$ and $\Omega_{\lambda}=0.7$. Middle: An open
universe with $\Omega_{m}=0.1$ and $\Omega_{\lambda}=0.5$. Right
panel: A closed
 universe with $\Omega_{m}=1.0$ and $\Omega_{\lambda}=2.0$. \label{5}}
\end{figure}

\begin{figure}[pb]
\centerline{\psfig{file=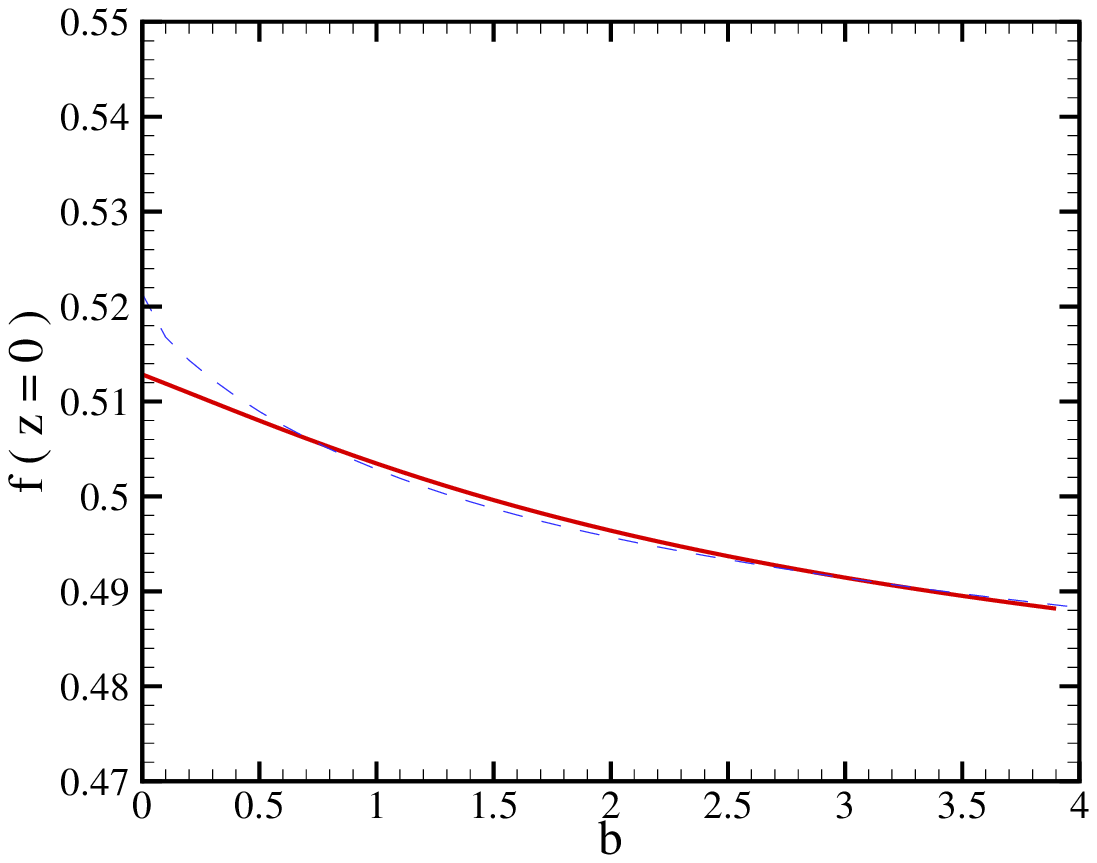,width=8cm}} \vspace*{8pt}
\caption{The growth index $f$ at the present time as a function of
the bending parameter $b$ for a flat universe with $\Omega_m=0.3$
and $\Omega_{\lambda}=0.7$. The solid line is given by the numerical
solution of Eq. (\ref{index}) and dashed-line is given by fitting
formula (Eq. (\ref{fit})). \label{6}}
\end{figure}

At the present time the growth  index shows mostly a dependence on
the matter density of the universe $\Omega_m$. But the time
evolution of the growth index depends strongly on the choice of the
parameters $\Omega_\lambda$ and $b$. Fig. \ref{5} shows the time
evolution of the growth index $f$ for different pairs of
$(\Omega_m,\Omega_{\lambda})$, where the sensitivity to the bending
parameter is examined. The main effect is that  increasing the
bending parameter decreases the growth index for all cases of open,
closed and flat universe and results in a lower abundance of
structure at the present time. In Fig. \ref{6} the result from
numerical solution of Eq. (\ref{index}) and from fitting formula,
Eq. (\ref{fit}), are illustrated. This shows that the growth index
is approximated by $f\sim \Omega_m^{\alpha}$.

In the case of a closed universe, the growth index rises to a
maximum and descends afterward. This temporal increasing of the
growth index is due to the dynamical effect of dark energy that
changes the sign of the acceleration from negative to positive and
the slope of deceleration parameter $q=-\ddot{a}/aH^2$. During the
transition phase of the acceleration, the universe reaches its
lowest expansion rate and the structures have an opportunity to grow
almost
exponentially. 
Increasing the bending parameter suppresses the bump of the growth
index (see Fig. \ref{5}), because the dominance of the dark energy
occurs at earlier times and increases the friction term proportional
to the expansion rate.

\section{Variable Dark energy and nonlinear structure formation in
the spherical approximation}

A simple approximation for calculating the formation of cosmological
objects is the spherical infall model \cite{LLPR91,GLKH03,piran}. In
this model we consider an overdense (or an underdense) spherical
region, i.e. a positive density perturbation $\delta>0$ (or
$\delta<0$) with spherical symmetry. For large enough positive
perturbations, this region becomes gravitationally unstable and
grows to a bound structure seen in the universe today. In such
overdense regions the gravity of the perturbations is able to stop
the expansion, turn it around and finally make the particles
collapse. This highly nonlinear process is usually modeled with a
number of simplifying assumptions \cite{P93}. Here we also apply
this approximation to the case of a negative density perturbation
$\delta<0$ and take into account a non-constant dark energy term.
For $\delta<0$ only the initial conditions are different and the
dynamical growth of the spherical underdense region is governed by
the same equation. Since in the underdense regions the probability
of forming galaxies, in particular luminous galaxies is lower,
\cite{GLKH03}, these regions correspond to voids in the
observed galaxy distribution. \\
We begin with the Friedmann equation with a quintessence term as
given in Eq. (\ref{eq4}). Until an initial time $t_i$ we assume the
spherical region $R_i$ to grow with the same rate as the background
universe: \beq
\left(\frac{\dot{R_i}}{R_i}\right)^2=H_0^2\left[\Omega_ma_i^{-3}
+\Omega_{\lambda}a_i^{-3[1+\bar{w}(a_i;b,w_0)]}-(\Omega_{tot}-1)a_i^{-2}
\right], \label{sphr3}, \eeq where $a_i$ is the initial scale factor
of the background. At the time $t_i$ we introduce an initial density
perturbation $\delta_i$ in the matter density
$\Omega_m^i=(1+\delta_i)\Omega_ma_i^{-3}$. As long as there is no
shell crossing, the matter content inside a shell is constant. As in
Section 2, we assume the effect of dark energy to be on much larger
scales than the forming structure. This means that the dynamics of
the universe is changed by dark energy but at the scale of
structures, such as galaxies and clusters of galaxies the pressure
of dark energy is negligible and the energy is conserved. Using the
energy conservation for the spherical region with a radius $R$ and a
mass $M$ yields:

\beq \dot{R}^2=2\left[E+\frac{GM}{R}\right] \label{sphr1} \eeq the
potential energy at every time is:

\beq
2\frac{GM}{R}=(H_0R)^2\left[\Omega_m(1+\delta)a^{-3}+\Omega_{\Lambda}(a;b,w_0)\right]
\label{sphr2}, \eeq using $M=M_i$, we get:

\beq
\left(\frac{R}{a}\right)^3(1+\delta)=\left(\frac{R_i}{a_i}\right)^3(1+\delta_i)
\label{sphr4}, \eeq therefore by using equations (\ref{sphr3}),
(\ref{sphr2}) and (\ref{sphr4}) we obtain for the size evolution of
the spherical region $R$ in the Eq. (\ref{sphr1}):

\begin{eqnarray}
\frac{\dot{R}^2}{H_0^2}=
\frac{\Omega_m(1+\delta_i)}{a_i^3}\frac{R_i^3}{R}
+\Omega_{\lambda}R^2a^{-3[1+\bar{w}(a;b,w_0)]}\nonumber\\
\qquad\qquad-\Omega_m\delta_i\frac{R_i^2}{a_i^3}
-(\Omega_{tot}-1)(\frac{R_i}{a_i})^2. \label{sphr51}
\end{eqnarray}

\begin{figure}[pb]
\centerline{\psfig{file=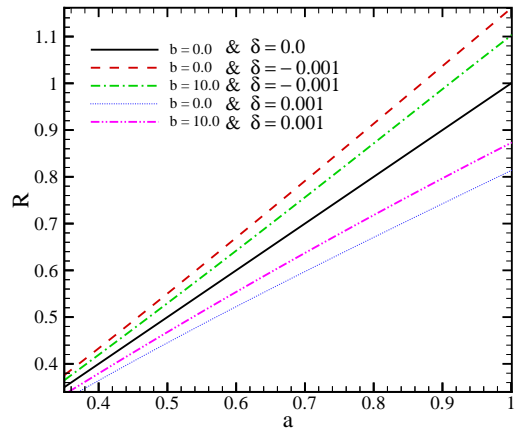,width=8cm}} \vspace*{8pt} \caption{
Evolution of radius  of a spherical overdense or underdense
perturbation as a function of scale factor for different bending
parameters. The initial conditions are $z_{i}=1100$ and $\delta_{i}$
as denoted in the diagram. \label{10}}
\end{figure}

After the time $t_i$ the scale factor $a$ evolves according to the
Friedmann equation and the spherical region changes its size with
Eq. (\ref{sphr51}). For the numerical solution of this equation we
replace $\dot{R}$ by ${dR}/{da}$. Fig. \ref{10} shows the evolution
of $R$ as a function of scale factor $a$. In agreement with the
results obtained by the linear perturbation theory (Fig. \ref{3}),
Section 2, we also see a suppression of the growth of overdense and
underdense regions if we increase $b$. Although the effect on the
dynamical growth of spherical underdense regions seems to be very
small, its tendency is discovered by observational data. It has been
shown for example in \cite{AM02,MAET00}, that large voids have a
lower abundance in the cosmological $N$-body simulations than in the
observations of the large scale in redshift surveys. The effect of a
variable dark energy  would slightly increase this discrepancy.

\section{Observational constraints}

In this section we examine the constraints on the free parameters of
the model from the observational data of the large scale structure
(LSS) and SNIa. Observational constraints from the CMB and the
amplitude of baryonic oscillations peak from the luminous red
galaxies in Sloan Digital Survey Sky (SDSS) can be found in
\cite{sadegh06}.

\subsection{Constraints from Supernovae Type Ia}
The SNIa experiments provide the most important indication for the
existence of dark energy in the standard model of cosmology. Since
1995 two teams, the High-Z Supernovae Search, and the Supernovae
Cosmology Project have discovered several SNIa candidates at high
redshifts \cite{Schmidt,per99}. In \cite{R04} the discovery of $16$
SNIa the Hubble Space Telescope were announced containing some of
the most distant ($z> 1.25$) SNIa known to date. Based on this data
a uniform Gold sample of high and low redshift SNIa was constructed,
\cite{R04,Tonry,bar04}. In this subsection we compare the
predictions of the dark energy model with the SNIa Gold sample. The
observations of Supernovae measure essentially the apparent
magnitude $m$ including reddening, K correction etc, which is
related to the (dimensionless) luminosity distance, $D_L$, of a an
object at redshift $z$, for a spatially flat universe by: \beq
m=\mathcal{M}+5\log{D_{L}(z;b,w_0)}, \label{m} \eeq where for
arbitrary spacial geometry
\begin{eqnarray}
\label{luminosity} D_L (z;b,w_0) &=&
\frac{H_0(1+z)}{\sqrt{|\Omega_K|}}{\rm
sinn}\left(\sqrt{|\Omega_K|}\int^{z}_{0}{1\over
H(\zeta;b,w_0)}d\zeta\right),
\end{eqnarray}
where $\Omega_K = 1 - \Omega_m - \Omega_{\Lambda}$, and "sinn" is
"sinh" for $k> 0$ (closed Universe) and "sin" for $k < 0$ (open
Universe) \cite{caroll92}. For k = 0, Eq. (\ref{luminosity}) reduces
to $H_0 (1 +z)$ times the integral. Also
\begin{eqnarray}
\label{m1}\mathcal{M} &=& M+5\log_{10}{\left(\frac{c/H_0}{1\quad
Mpc}\right)}+25,
\end{eqnarray}
where $M$ is the absolute magnitude. The distance modulus, $\mu$, is
defined as:

\beq \mu\equiv
m-M=5\log_{10}{D_{L}(z;b,w_0)}+5\log_{10}{\left(\frac{c/H_0}{1\quad
Mpc}\right)}+25, \label{eq:mMr} \eeq

In order to compare the theoretical results with the observational
data, we must compute the distance modulus, as given by Eq.
(\ref{eq:mMr}). The first step in this sense is to compute the
quality of the fitting through the least squared fitting quantity
$\chi^2$ defined by:
\begin{eqnarray}\label{chi_sn}
\chi^2=\sum_{i}\frac{[\mu_{obs}(z_i)-\mu_{th}(z_i;\Omega_m,w_0,b,h)]^2}{\sigma_i^2},
\end{eqnarray}
where $\sigma_i$ is the observational uncertainty in the distance
modulus. To constrain the parameters of model, we use the Likelihood
statistical analysis. The method and its motivation are described in
detail in ref. \cite{co04}. In the absence of prior constraint, the
probability of the set of distance modulus $\mu$ conditional on the
values of a set of model parameters is given by a product of
Gaussian functions: \beq
p(\mu_{th}(z_i;\{l\}))=\prod_i\frac{1}{\sqrt{2\pi\sigma_i^2
}}\exp\left[-\frac{[\mu_{obs}(z_i)-\mu_{th}(z_i;\{l\})]^2}{2\sigma_i^2}\right],
\eeq where $\{l\}=\{\Omega_m,w_0,b,h\}$. This probability
distribution must be normalized. Evidently, when, for a set of
values of the parameters, the $\chi^2$ is minimum the probability is
maximum. We find the minimum value of $\chi^2_{min}/N_{d.o.f}=1.131$
corresponding to the best fit values for the model parameters
$h=0.66$, $w_0=-1.90_{-3.29}^{+0.75}$,
$\Omega_m=0.01^{+0.51}_{-0.01}$ and $b=6.00_{-6.00}^{+7.35}$ at
$1\sigma$ level of confidence. As a special case we fix the value of
the state parameter to $w_0=-1.0$ and obtain the best fit values for
the model parameters as: $h=0.65$, $\Omega_m=0.01^{+0.14}_{-0.01}$
and $b=2.08_{-0.98}^{+0.40}$ at $1\sigma$ confidence level with
$\chi^2_{min}/N_{d.o.f}=1.133$. The values of cosmological
parameters from fitting the dark energy model are different from
those in the $\Lambda$CDM model, i.e. the value of $h=
0.71_{-0.03}^{+0.04}$ and $\Omega_m=0.27^{+0.04}_{-0.04}$ are
slightly smaller in the WMAP results for $\Lambda$CDM model
\cite{spe03,eidelman04}. If we use the Hubble parameter
$H_0=71.0\pm7.0$ from the HST-Key project as a prior parameter, the
best fit values are: $\Omega_m=0.00^{+0.01}_{-0.00}$ and
$b=0.79_{-0.13}^{+0.13}$ at $1\sigma$ confidence level with
$\chi^2_{min}/N_{d.o.f}=1.552$. The values of model parameters with
$1\sigma$ and $2\sigma$ confidence levels are summarized in Table
\ref{tab1} and \ref{tab2}. Figures \ref{mbhw}, \ref{mbh} and
\ref{mb_likelihood_hst} show marginalized relative probability
density functions for cosmological parameters.

\subsection{Combined Constraints from Structure formation and SNIa}
The above analysis shows that the SNIa data alone does not
sufficiently constrain the variable dark energy model. Furthermore
the fit of SNIa data is sensitive to the parameter $H_0$. Hence, it
is very important to find other observational quantities independent
of $H_0$ as a complement to the SNIa data. In the pervious sections
we studied the effect of varying dark energy model on the evolution
of large scale structures. Here we use the results from the $2$dFGRS
data and combine this with the SNIa data to put more rigorous
constraint on the parameters of variable dark energy model.

The 2dFGRS contains the observation of the position and the redshift
of about $220,000$ galaxies. On this basis the 2dFGRS team has
determined a growth index $f$, which is the relevant parameter for
the comparison with the predictions of the dark energy model
\cite{eisenstein05}. From measuring the two-point correlation
function, they report the redshift distortion parameter $\beta =
{f}/{\tilde{b}} =0.49\pm0.09$ at $z=0.15$, where $\tilde{b}$ is the
bias parameter describing the difference in the distribution of
galaxies and mass. Verde et al. (2003) used the bispectrum of
$2$dFGRS galaxies \cite{ver01,la02} and obtained $\tilde{b}_{verde}
=1.04\pm 0.11$, from which, we obtain $f(z=0.15)= 0.51\pm0.10$.

\begin{figure}[pb]
\centerline{\psfig{file=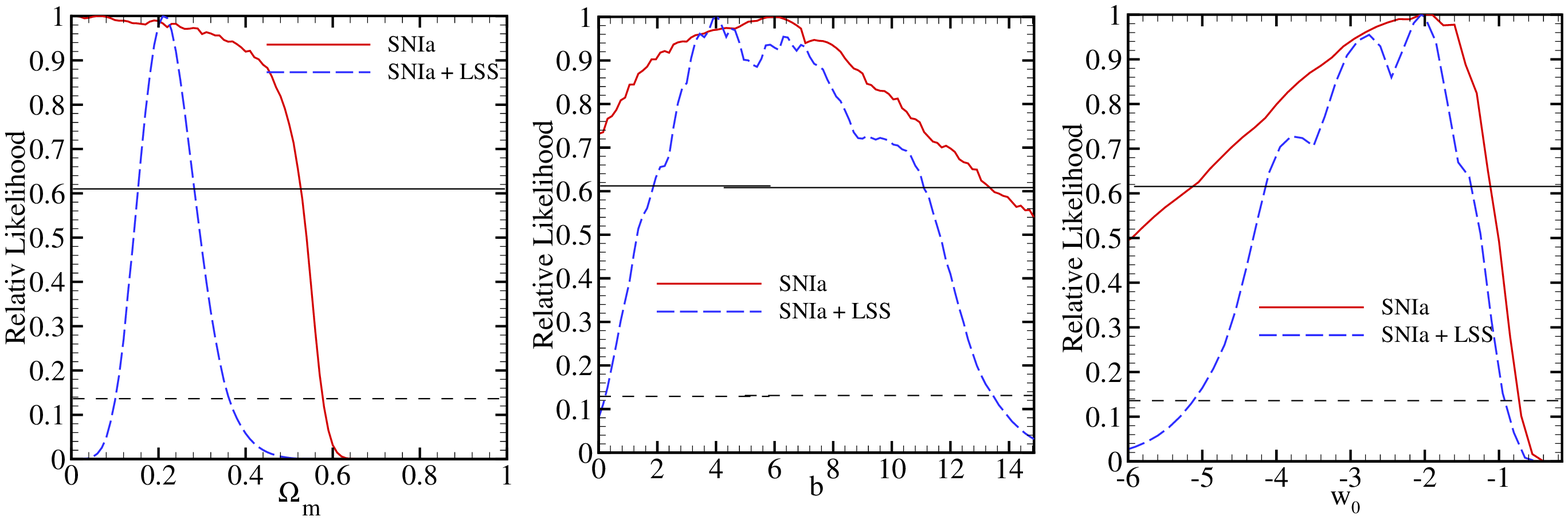,width=15cm}}
\centerline{\psfig{file=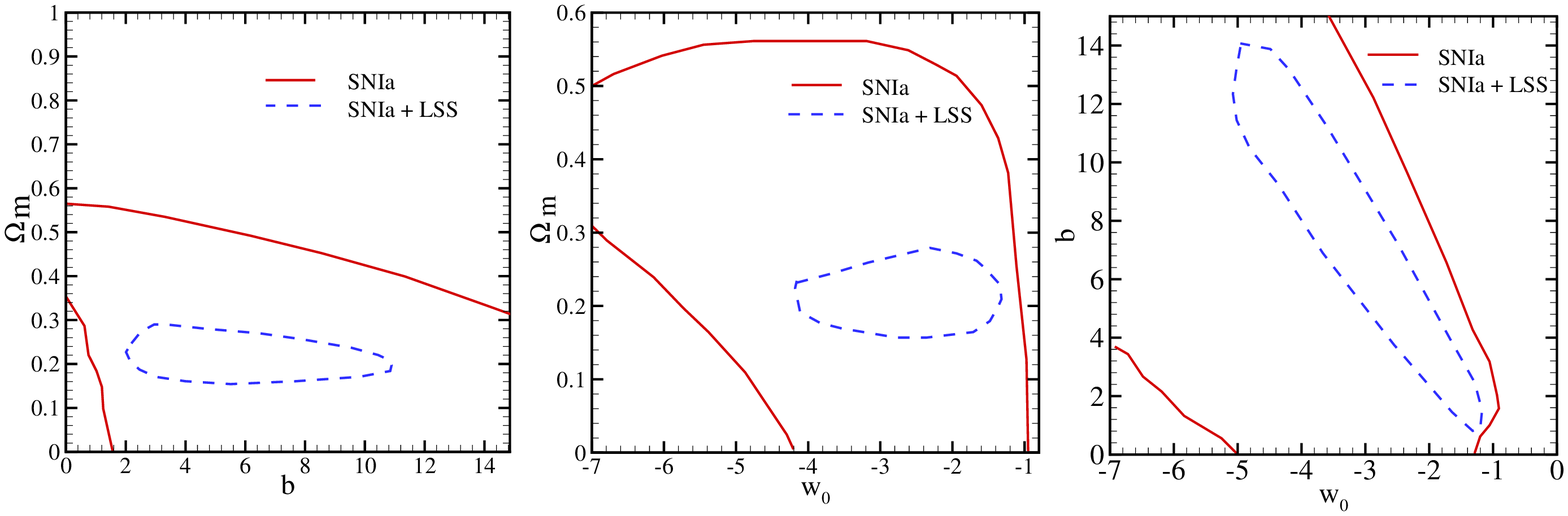,width=15cm}}  \vspace*{8pt}
\caption{The upper panel shows marginalized likelihood functions of
cosmological parameters. The solid line was obtained by
marginalizing over all nuisance parameters, using SNIa and
dashed-line corresponds to SNIa$+$LSS. The intersections with the
horizontal solid and dashed lines give the bounds for $68.3\%$ and
$95.4\%$ confidence respectively. The lower panel shows joint
confidence contours with $1\sigma$ confidence level. \label{mbhw}}
\end{figure}

We perform a combined analysis of the SNIa and Large Scale Structure
(LSS) to constrain of variable dark energy model through $\chi^2$
fitting:
\begin{equation}
\chi^2=\chi^2_{\rm {SNIa}}+\chi^2_{{\rm LSS}},
\end{equation}
where $\chi^2_{{\rm SNIa}}$ is given by Eq. (\ref{chi_sn}) for SNIa
data and $\chi^2_{{\rm LSS}}$ is the contribution of LSS data using
Eq. (\ref{index}). The best fit values for the model parameters are:
 $h=0.66$, $\Omega_m=0.21^{+0.07}_{-0.06}$, $b=4.05_{-2.25}^{+7.05}$
and $w_0=-2.05_{-2.05}^{+0.65}$ at $1\sigma$ confidence level with
the corresponding $\chi^2_{min}/N_{d.o.f}=1.131$. The values of
model parameters for $1\sigma$ and $2\sigma$ confidence levels using
SNIa$+$LSS observations are given in Table \ref{tab1}. Fig.
\ref{mbhw} shows the marginalized relative likelihood functions
(upper panel) and joint confidence contours (lower panel) of the
model parameters.

\begin{figure}[pb]
\centerline{\psfig{file=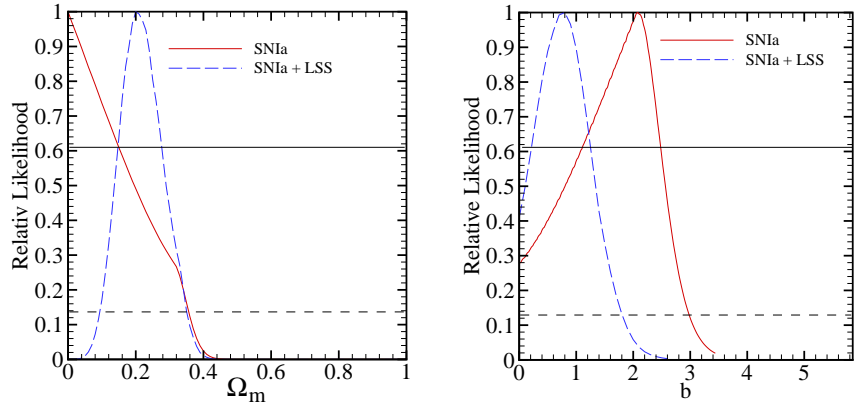,width=12cm}} \vspace*{8pt}
\caption{The marginalized likelihood functions of cosmological
parameters. The solid line was obtained by marginalizing over all
nuisance parameters, using SNIa and dashed-line corresponds to
SNIa$+$LSS. The intersections with the horizontal solid and dashed
lines give
the bounds for $68.3\%$ and $95.4\%$ confidence respectively. 
Here we assume $w_0=-1.0$. \label{mbh}}
\end{figure}

\begin{figure}[pb]
\centerline{\psfig{file=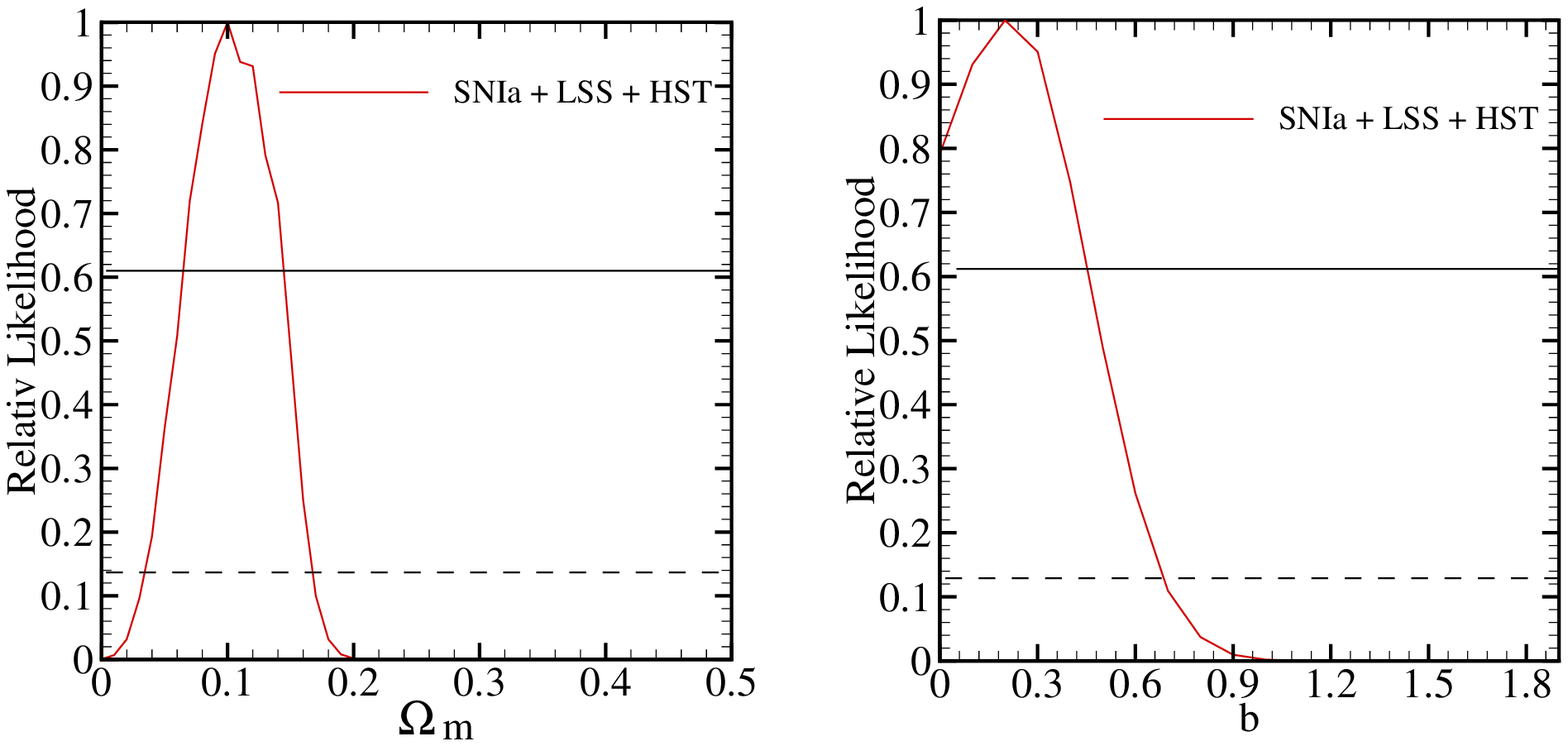,width=12cm}} \vspace*{8pt}
\caption{Marginalized likelihood functions of cosmological
parameters. The solid line was obtained by marginalizing over all
nuisance parameters, using SNIa$+$LSS$+$HST. HST prior is
$H_0=71.0\pm7.0$. The intersections with the horizontal solid and
dashed lines give the bounds for $68.3\%$ and $95.4\%$ confidence
respectively. Here we assume $w_0=-1.0$. \label{mb_likelihood_hst}}
\end{figure}

Further, we restrict the analysis to a subset of the parameter space
and fix the present state parameter at $w_0=-1.0$. Again using the
combined LSS and SNIa data, the best fit values for the model
parameters are $h=0.64$, $\Omega_m=0.20^{+0.07}_{-0.05}$ and
$b=0.80_{-0.60}^{+0.45}$ with $\chi^2_{min}/N_{d.o.f}=1.142$ at
$1\sigma$ confidence level. The values of the model parameters for
$1\sigma$ and $2\sigma$ confidence levels are reported in Table
\ref{tab2}. Fig. \ref{mbh} shows the marginalized relative
likelihood functions of matter density and the bending parameter. If
we use $H_0=71.0\pm7.0$ from the HST-Key project as a prior with $1
\sigma$ measurement, we obtain the best fit values of the model
parameters at $\Omega_m=0.10_{-0.03}^{+0.04}$,
$b=0.20_{-0.20}^{+0.25}$ at $1\sigma$ confidence level with
$\chi^2_{min}/N_{d.o.f}=1.618$. The marginalized relative likelihood
functions of the matter density and the bending parameter are shown
in Fig. \ref{mb_likelihood_hst}.

Finally we determine the predicted age of universe considering the
parameters of the model according to:
\begin{equation}\label{age}
t_0(b,w_0) = \int_0^{t_0}\,dt = \int_0^\infty {dz\over
(1+z)H(z;b,w_0)},
\end{equation}
 and compare it with the age of old objects in the universe as a consistency
test. In Table \ref{tab1} we show that the age of universe from SNIa
and the combined SNIa$+$LSS analysis are $13.45^{+3.39}$Gyr and
$13.54^{+2.34}_{-3.75}$ Gyr, respectively, which is in the range of
the age of old stars $13^{+4}_{-2}$ Gyr \cite{carretta00}.

Furthermore we use the observed age of three Old High Redshift
Galaxies (OHRG) for comparison with the dark energy model, namely
the LBDS $53$W$091$, a $3.5$-Gyr-old radio galaxy at $z=1.55$
\cite{dunlop96}, the LBDS $53$W$069$, a $4.0$-Gyr-old radio galaxy
at $z=1.43$ \cite{dunlop99} and a quasar, APM $08279+5255$ at
$z=3.91$ with an age of $t=2.1_{-0.1}^{+0.9}$Gyr \cite{hasinger02}.
The latter one has once led to the "age crisis" in the $\Lambda$CDM
model \cite{jan06}. To quantify the age consistency test we
introduce the expression $\tau$ as:
\begin{equation}
 \tau=\frac{t(z;b,w_0)}{t_{obs}} ,
\end{equation}
where $t(z;b,w_0)$ is the predicted age of universe from Eq.
(\ref{age}) and $t_{obs}$ is an estimation for the age of the old
object. In order to have a compatible predicted age of the universe,
it is necessary to have $\tau\geq1$. Table \ref{tab3} shows the
values of $\tau$ for three mentioned OHRG. If we fix the parameters
within the 1-$\sigma$ confidence level of the above combined
observational constraints of SNIa and LSS, the predicted age of the
universe is larger than the age of LBDS $53$W$069$ and LBDS
$53$W$091$, while APM $08279+5255$ at $z=3.91$ is older. Only in the
case that we fix $w_0=-1$ and $H_0=71.0\pm7.0$ from HST-Key project,
we obtain $\tau>1$.

\begin{table}[ph]
\tbl{Estimated variable dark energy model parameters (mean,
$68.3\%$, $95.4\%$ C.L.), including the present matter density, the
bending parameter, the state parameter and the age of Universe.}
{\begin{tabular}{@{}ccccc@{}} \toprule  Observation & $\Omega_m$& $b$& $w_0$& age (Gyr)\\
 \colrule
 SNIa & $0.01^{+0.51}_{-0.01}$&$6.00^{+7.35}_{-6.00}$&$-1.90^{+0.75}_{-3.29}$ &$13.45^{+3.39}_{-13.45}$  \\ 
  & $0.01^{+0.56}_{-0.01}$&$6.00^{+17.42}_{-6.00}$&$-1.90^{+1.10}_{-7.23}$ &
  \\&&&&\\
   SNIa+LSS & $0.21^{+0.07}_{-0.06}$&$4.05^{+7.05}_{-2.25}$&$-2.05^{+0.65}_{-2.05}$& $13.54^{+2.34}_{-3.74 }$ \\
     &$0.21^{+0.17}_{-0.11}$&$4.05^{+9.20}_{-3.75}$&$-2.05^{+1.10}_{-3.10}$&
\\ \botrule
\end{tabular} \label{tab1}}
\end{table}

\begin{table}[ph]
\tbl{Estimated variable dark energy model parameters (mean,
$68.3\%$, $95.4\%$ C.L.) for the fixed value of the state parameter
$w_0=-1$, including the present matter density, the bending
parameter and the age of Universe assuming $w_0=-1.0$. The HST prior
is set at $0.71\pm0.07$.}
{\begin{tabular}{@{}cccc@{}} \toprule  Observation & $\Omega_m$& $b$& age (Gyr)\\
 \colrule
  SNIa & $0.01_{-0.01}^{+0.15}$&$2.08^{+0.40}_{-0.98}$&$14.23^{+1.27}_{-2.20}$   \\ 
 &
$0.01_{-0.01}^{+0.35}$&$2.08_{-2.08}^{+0.89}$&
\\&&&\\
  SNIa+HST & $0.00_{-0.00}^{+0.01}$&$0.79^{+0.13}_{-0.13}$ &$16.66^{+0.64}_{-1.09}$   \\ 
 &
$0.00_{-0.00}^{+0.06}$&$0.79^{+0.28}_{-0.36}$ &
\\&&&\\
SNIa+LSS
&$0.10^{+0.04}_{-0.03}$&$0.20^{+0.25}_{-0.20}$&$16.72^{+1.52}_{-1.43}$ \\
+HST&$0.10^{+0.06}_{-0.07}$&$0.20^{+0.45}_{-0.20}$&\\&&&\\ SNIa+LSS
& $0.20_{-0.05}^{+0.07}$&$0.80^{+0.46}_{-0.60}$&$14.55^{+1.04}_{-1.44}$ \\
&$0.20_{-0.10}^{+0.15}$&$0.80^{+0.90}_{-0.80}$&\\
 \botrule
\end{tabular} \label{tab2}}
\end{table}

\begin{table}[ph]
\tbl{Estimated relative age $\tau$ of three high redshift objects
using the best fit values of model parameters with $1\sigma$ confidence level.} {\begin{tabular}{@{}cccc@{}}
\toprule  Observation & LBDS $53$W$069$&LBDS $53$W$091$& APM $08279+5255$ \\
  & $z=1.43$&$z=1.55$& $z=3.91$ \\
  \colrule \\
SNIa$+$$w_0=-1$& $1.13^{+0.31}_{-0.55}$ & $1.21^{+0.36}_{-0.63}$& $0.78^{+0.69}_{-0.78}$ \\
&&& \\

 SNIa+HST+$w_0=-1$ & $1.67^{+0.16}_{-0.27}$&$1.81^{+0.18}_{-0.31}$&$1.26^{+1.10}_{-0.52}$ \\
&&&\\

 SNIa+LSS+$w_0=-1$ & $1.19^{+0.26}_{-0.36}$&$1.27^{+0.30}_{-0.41}$&$0.84^{+0.61}_{-0.68}$ \\
  &&&\\
SNIa+LSS+HST+$w_0=-1$ & $1.63^{+0.38}_{-0.35}$&$1.75^{+0.43}_{-0.41}$&$1.19^{+0.88}_{-0.68}$ \\
&&&\\

 SNIa & $1.00^{+0.84}_{-1.00}$&$1.05^{+0.96}_{-1.05}$&$0.65^{+1.63}_{-0.65}$
 \\&&&\\
  SNIa+LSS & $1.00^{+0.58}_{-0.93}$&$1.06^{+0.67}_{-1.06}$&$0.67^{+1.15}_{-0.67}$ \\
 \botrule
\end{tabular} \label{tab3}}
\end{table}

\section{Conclusions}
In this paper we have considered a simple model for variable dark
energy and examined its influence on the cosmological structure
formation. First we have used the Newtonian linear perturbation
theory to see the possible effects of a quintessence model, which is
parameterized according to a prescription given by Wetterich (2004).
The crucial parameter of this scheme is the bending parameter $b$,
which changes the behavior of dark energy toward a more dust-like
behavior. Considering a flat, an open and a closed universe and an
almost arbitrary initial value of the density contrast $\delta$, we
find that by increasing the bending parameter, the growth of the
$\delta$ is decreased in comparison to the conventional $\Lambda$CDM
model. In agreement with the results of our nonlinear spherical
calculations, we find that increasing the bending parameter reduces
the rate of growth of structure. This suppression in the growth of
structure occurs in all open, flat and closed models for
sufficiently large values of $b$.

This effect can be explained by the fact that the dominance of the
dark energy term occurs earlier with respect to $\Lambda$CDM, the
larger the parameter $b$ is. Because of the earlier domination of
dark energy, the rate of the structure formation process is
decreased. This finding is also supported by the second part of our
calculations, where we numerically solve the equation of the
collapse of a spherically symmetric density perturbation. We also
give the fitting formula for the growth index, $f$, at the present
time as a function of model parameters. the leading term of this
formula is $\Omega_m^{\alpha}$ which is in agreement with
$\Lambda$CDM models \cite{sil,wang98,lokas}.

In order to constrain the model parameters, we used the Gold sample
SNIa data. The SNIa analysis resulted in a large degeneracy between
the model parameters. For improving this, we perform a joint
analysis of the SNIa and LSS data. This joint analysis of SNIa$+$LSS
produces more reasonable results: $h=0.66$,
$\Omega_m=0.21_{-0.06}^{+0.07}$, $w_0=-2.05_{-2.05}^{+0.65}$ and
$b=4.05_{-2.25}^{+7.05}$, at $1\sigma$ confidence level with
$\chi^2_{min}/N_{d.o.f}=1.131$. The age of universe from the best
fit values of model parameters by using SNIa$+$LSS is
$13.54^{+2.34}_{-3.74}$Gyr, which is compatible with the age of old
stars. We also chose three high redshift radio galaxies namely the
LBDS $53$W$091$, a $3.5$-Gyr-old radio galaxy at $z=1.55$, the LBDS
$53$W$069$, a $4.0$-Gyr-old radio galaxy at $z=1.43$ and a quasar,
APM $08279+5255$ at $z=3.91$ with an age of
$t=2.1_{-0.1}^{+0.9}$Gyr. The age of the two first objects is
consistent with the age of universe using the best fit values of the
model parameters, while the later one is older than the universe.
The age of quasar APM $08279+5255$, was also the reason of the "age
crisis" for the $\Lambda$CDM model and it is compatible with the
dark energy model only if we fix $w_0=-1.0$ and assume the Hubble
parameter at the value given by HST-Key project as a prior in our
calculations.

\section*{Acknowledgments}
The authors thank Niyayesh Afshordi and Volker M\"uller for their
useful comments.

\end{document}